# Real Options Technique as a Tool of Strategic Risk Management

*Vladimir Savchuk*
*International Institute of Business, Kiev*
*Royal Holloway University of London*

**Abstract**

The real options approach is now considered an effective alternative to the corporate DCF model for a feasibility study. The current paper offers a practical methodology employing binomial trees and real options techniques for evaluating investment projects. A general computation procedure is suggested for the decision tree with two active stages of real options, which correspond to additional investments. The suggested technique can be used for most real options, which are practically essential regarding enterprise strategy. The special case named Binomial-Random-Cash-Flow Real Options Model with random outcomes is developed as the next step of real options modelling. "Project Value at Risk" is introduced and used as a criterion of investment project feasibility under the assumption regarding random outcomes. In particular, the Gaussian probability distribution is used for modelling option outcomes uncertainty. The choice of the Gaussian distribution is caused by the desire to obtain estimates in the final analytical form. Choosing another distribution for random outcomes leads to using Monte Carlo simulation, for which a general framework is developed by demonstrating some instances. The author could avoid the computational complexity that makes these solutions feasible for business practice.

**Introduction**

From the practical standpoint, the Real Options approach deals with the strategic justification of a firm's investment projects. All investment projects aim to receive an expected return from the investment. The risks of obtaining the expected returns are becoming increasingly high due to the high level of uncertainty regarding the economic and political situation in modern society. In such a situation, traditional methods of Capital Budgeting, the so-called DCF approach, based on discounting cash flows and calculating NPV/IRR metrics, can hardly be considered adequate. A real option represents a way to improve the DCF approach, which allows for more effectively assessing and managing real assets. As a result of this, the real options approach can be considered a conceptual tool for making strategic decisions and developing powerful strategies. It should be stressed that enterprise managers often use the real option phenomenon intuitively and need to formalize their intuitive suggestions with quantitative estimates.

In actual practice, the term "decision-making" implies the choice of a specific decision and its subsequent implementation. At the same time, there are often situations when the decision-maker, having made a decision, still determines if its implementation should be started immediately or whether it can be possible to change what is essential for the company for now. Investment decisions can be considered a typical example of such a situation. These decisions involve a long implementation period, which increases the degree of uncertainty and, as a result, increases the risks of not receiving the expected return or even loss of invested funds.

Real options have been a popular topic in financial literacy over the last two decades. Increasingly, the discussion of real options, methods of their analysis, and their integration into strategic planning procedures can be found in specialized financial publications, books, and articles on management. Nevertheless, real options analysis is often considered excessively complicated or unrelated to real life. In fact, it's precisely the opposite. Valuation of real options helps to mitigate the significant shortcomings of traditional methods of financial calculations and apply quantitative approaches to investment decisions where these methods do not work. This means the possibility of a more adequate analysis of strategic decisions.

The real option, as a definition, was introduced by Stewart Myers (1977). He tried to apply the theory of financial options traded in the derivatives market to analyse a company's financial policy using financial leverage. The real option Myers considered concerned a firm's financial policy using leverage. According to his interpretation, shareholders own a call option on the right to own a firm. They will abandon the firm if its value is insufficient, transferring ownership rights to creditors. This approach allows us to assess the market value of debts.

More recently, this approach has been used to assess opportunities that result from strategic decisions that may arise in the future. The main difference between real and financial options is that a real option is not a security. It does not circulate in the derivatives market, where it can be sold or bought. The underlying asset of a real option is the future management decisions that can be made in relation to a specific development project.

Brennan and Schwartz (1985) developed the idea of real options by applying it to the valuation of natural resources (oil and copper mining, etc). Another pioneering work of McDonald and Siegel (1986) was devoted to the economic assessment of time as a strategic resource meaning the possibility of postponing the start of investments.

An essential stage in forming the theory of real options was the publication of a monograph by Dixit, and Pindyck (1994), which for the first time, considered the general analytical theory within the financial economics framework taking into account the position of the company in the market. In the late 1990s, there are examples of actual practical applications of real options. One of the first such cases at Merck was documented by Nichols (1994). A variety of applications of real options theory at Intel was done by Sammer (2002).

The main monographic works on real options are Amram, Kulatilaka (1999); Brennan, Trigeorgis, (2000); Copeland, Antikarov, (2001); Dixit, Pindyck (1994), Trigeorgis (1995, 1996).

Researching the phenomenon of real options corresponds to the general trends in the economy and business development associated with the aggravation of competition due to globalisation and the accelerated growth of technological progress. In addition, even in traditional industries, companies in transition economies face an unpredictable external environment in which it is impossible to foresee not only the actions of competitors (the cultural environment of business is not formed), but also the regulatory steps of the authorities (as well as their behaviour). What has been said here is directly related to Ukraine. The author of this paper was dealing with some Ukrainian investment projects which didn't demonstrate their feasibility by the traditional DCF approach. Some of them turned out to be attractive due to the assumption of possible real options

From the point of view of the actual practice of strategic management, real options can be interpreted from two positions. The first, light position, involves using real options techniques to solve various problems, primarily related to Capital Budgeting. In this case, real options are treated as an analogue of financial options, which implies the transfer of the theory of financial options to real assets. Since such a theory is quite complex for practitioners, there is a risk that it will primarily remain a theory and not reach wide distribution in practice. The second, more comprehensive, position is that real options should create the basis of the company's development strategy. In this case, they are treated as a universal management process, a phenomenon of strategy in dynamics. The company's management should see and evaluate the options when developing a strategy. And this creates a real competitive advantage for the firm. We lose alternative profits and incur additional expenses if we do not see real options.

Both interpretations allow us to determine the real options toolkit for making flexible decisions under uncertainty. The methods considered in this paper tend to the second, more rigid understanding of the phenomenon of real options. The author attempts to minimise the risk of being misunderstood from the point of view of the practitioner decision-maker.

One more essential point should be considered regarding the real options approach. This is a company's strategic risk. Uncertainty is an attributive source of risk. A priori, the risk is directly dependent on uncertainty: as uncertainty increases, so does risk. The magnitude of the increase may vary due to the elasticity of risks concerning uncertainty.

Two approaches dominate management to understanding risk. The first approach understands risk only as losses and dangers that await the company. In Merriam-Webster's Collegiate Dictionary, we will find a definition of risk as the "possibility of loss or injury", and to take a risk means "to expose to hazard or danger". This is how most people understand it, i.e. risk is determined mainly negatively. But this is correct only in relation to natural and man-made disasters, but not in the case of economics and management. The keywords in this approach are stability and security. This is not what management does. Management should ensure growth associated with the maximum increase in its value. Stability in a competitive environment always means stagnation which leads to crisis. A guaranteed result is the opposite of growth. Real options are strategic management of the growth process. In confirmation, we point to the recognition of such a prominent management theorist as Mintzberg (2001), who writes that risk was understood in management solely as a danger in the early times. Still, with the advent of real options, the situation has changed.

We can formalize such a situation by means of the error matrix presented in Fig. 1.1.

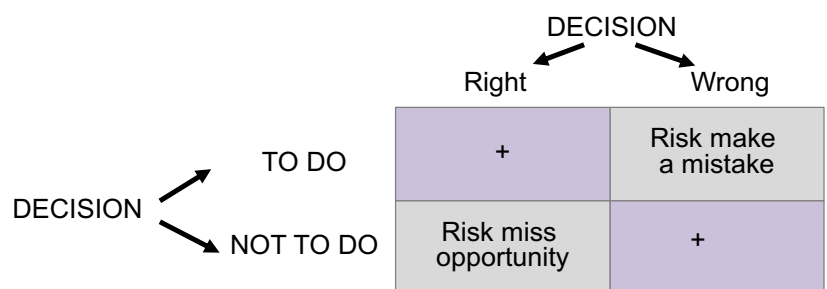

Fig. 1.1 Project Error Matrix

According to the scheme, there are two risks: a) rejecting a project if it is correct and b) accepting a project if it is wrong. Every person faces such risks. Most people are afraid to make the wrong decision. Fewer people experience the risk of missing out on an opportunity to become more affluent. In business, it's more complicated than that. You can't look at the opportunities (good) and the threats (harmful) separately. If you focus only on threats without considering the positive opportunities, the business must be closed immediately. The fight against threats cannot be the goal of the company. The company exists in the hope of success but knows there are failures. Flexibility and real options help to deal with possible failures, but the struggle is meaningful only if you believe that success covers the costs of dealing with dangers. Analysis of positive exits from adverse situations is one of the aspects of designing real options. Note that this is not the only side – when considering growth options, there is a cascade of sequential options, which creates opportunities for implementing the next. The paper suggests a classification of the real options, each of which can be considered a stage of a company's strategic development.

A real option is a means of active management aimed at maximizing the company's value, unlike a financial option is a means of hedging risk. This means that for a fee (equal to the option's price), the holder of a financial option "cuts off" unfavourable opportunities. Full hedging is usually impossible in management, and flexibility (option price) is very expensive. This qualitatively changes the situation compared to financial options, where the price of the option is small compared to the price of the asset. This is due, in particular, to the fact that a financial option is an arbitrage-free contract between the parties in a competitive and liquid market (options and, most importantly - the underlying asset), whereas real options usually exist only in the head of the manager, and the second player is the environment of nature.

So, finally, there are two approaches to risk management. The passive approach involves reducing the possibilities of threats through derivative financial instruments and internal management standards. An active approach uses real options, fundamentally increasing the company's competitive advantages.

It should be mentioned that applications of these decision-making rules in management rest on simplifying assumptions, making their direct use problematic. Even after these assumptions, the models remain complex mathematically and, therefore, oftentimes unrealistic for implementation in management practice, which has not yet mastered even classical decision-making methods. Farragher, Kleiman, Sahu, (2001) and Graham and Harvey, (2002) demonstrate the capital budgeting decision-making practices in US companies, which look not too optimistic regarding the actual uses of the real options approach. Ragozzino R., Reuer J. Trigeorgis L. (2016) have argued that financial economics and enterprise strategy have worked in quasi-independent ways over the years, and neither has managed to do a holistic job independently. This gap can be overcome by developing the real options techniques, which correspond to the main principles of the real options theory and can be incorporated into enterprise strategic management.

The current paper suggests models that are simple mathematically and logically fit into the generally accepted capital budgeting process. Building models and calculation procedures, the author looked at them through the eyes of a financial manager, whose responsibility is to justify the company's development projects from the standpoint of economic criteria.

In using real options, the most common is the binomial model based on the binary decision tree. However, its application assumes assigning deterministic values to the outcomes of various options. Under increasing uncertainty, this assumption limits the scope of application since the decision maker is ready to model the outcomes of options using a probabilistic distribution. In the simplest case, he/she can assign an uncertainty interval, assuming a uniform distribution, or set the mean value and the standard deviation. In the latter case, the uncertainty of each individual outcome can be modelled by the Gaussian probability distribution.

The paper will demonstrate how the real option binomial model works. First, we present the binomial model's general mathematics and illustrate its application by the most straightforward real-option investment projects. Then, we discuss the possibility of generalization to the case of uncertain outcomes, which brings us to the Binomial-Random-Cash-Flow Option Model. For this model, we discuss obtaining analytical solutions for Gaussian random outcomes and Monte Carlo simulation estimation for any other probability distribution.

# 1. Why the DCF Approach is needed to be improved

Like any method in finance, the DCF approach is based on some simplifying assumptions. Firstly, the final decision regarding the feasibility is based on forecasting future cash flows. The result is correct as much as the assumptions in the calculation about the future prices, sale volumes and other parameters. The calculation of NPV is based on the most likely forecast of the development of events based on the information available to date. The second essential assumption of the DCF method is that the decision is made "here and now". Evaluation models do not consider the possibilities of adaptation to changing conditions (i.e., decision-making in the future). In actual practice, many strategic decisions companies implement despite negative NPV estimations. This results from profound management confidence, which cannot be substantiated numerically. Typical examples are R&D investments, implementing pilot projects to introduce new products, entering new markets or introducing new technologies. The third important feature of the DCF approach is assigning a discount rate. Risk is usually estimated from industry data. But investment projects of different enterprises (even belonging to the same industry) may contain different opportunities for adaptation, which means that the level of risk in them is not the same; thereby, the discount rate should also differ.

Because of the above, a diametrically opposite point of view regarding the analysis of investment decisions enters the arena: the strategy cannot be calculated quantitatively. Oftentimes we can hear from managers the following arguments in favour of projects with a negative NPV: " while this project is not financially profitable, it is important from the point of view of implementing our strategy"; "this decision strengthens our long-term competitive advantages", "this project creates new opportunities". The point is that the DCF approach cannot consider the economic environment's high level of dynamic variability. This is mainly because assumptions about the degree of uncertainty made before the start of the project can change significantly during its implementation.

Let's first consider how the DCF approach works for a binomial scheme. We will use the notion of "Project Value" *(PV)* as the present value of all projected free cash flows (CF) during the period of project implementation. Assuming the two scenarios, we have two project values:

$$V_0^{(i)} = \sum_{k=1}^{n} \frac{CF_k^{(i)}}{(1+r)^k}, i = 1,2, \qquad (1.1)$$

where $V_0^{(i)}$ is a Project Value for the $i$-th scenario, $CF_k^{(i)}$ - Cash Flow for $k$-th year ($k = 1,2,...n$) for $i$-th scenario.

Fig. 1.1 represents this scheme by Cash Flows, where $p^{(i)}$ is a probability of $i$-th scenario, $p^{(2)} = 1 - p^{(1)}$, and $V_0$ is the project value for time $t = 0$.

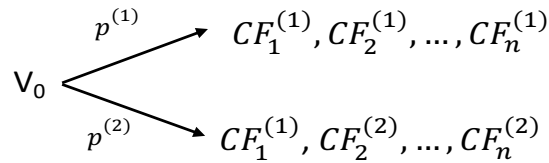

Fig. 1.1 DCF Value Flows for Binomial Model

To estimate Project Value for the $t = 0$, we need to find the expected value of the two scenarios.

$$V_0 = p^{(1)} \cdot V_0^{(1)} + p^{(2)} \cdot V_0^{(2)}. \qquad (1.2)$$

For the further modelling we will use the notion $V_k^{(i)}$ ($k = 1, 2, \ldots n$) of Project Value for all years of the Project. In particular

$$V_1^{(i)} = \sum_{k=2}^{n} \frac{CF_k^{(i)}}{(1+r)^{k-1}}, i = 1,2, \qquad (1.3)$$

Then the project Value for t = 0

$$V_0 = \frac{1}{(1+r)} \sum_{i=1}^{2} p^{(i)} \cdot \left(V_1^{(i)} + CF_1^{(i)}\right). \qquad (1.4)$$

It is easy to see that equations (1. 2) and (1.4) lead to the same result.

The DCF approach uses NPV as the most crucial metric for final decision-making, which is presented as a difference between initial investment $I_0$ and the project value.

$$NPV = -I_0 + V_0. \qquad (1.3)$$

What are the main strengths and weaknesses of the DCF approach?

The main advantage is the simplicity of calculating and interpreting the NPV metric. It is generally accepted that a project should be accepted if the NPV has a positive value. The number of this positive value is not discussed. And this immediately reveals the weakness of the method. After all, two positive numbers of $ 500K and $ 50K are very different. Still, simultaneously, two projects with such NPV values with an investment of $3,000K will be equally accepted for implementation. The second major weakness of the DCF approach is its very little risk concern. The scheme discussed above makes the first approach to taking into account risks - it considers two possible scenarios, and the number of scenarios may be more than two. At the same time, it does not allow us to consider the risks associated with the uncertainty of projected cash flows. Although it is this uncertainty that creates risks in an investment project.

A more advanced interpretation of NPV suggests that we consider the NPV metric as a margin of safety or strength for deciding to favour the investment project. The greater the value of the NPV, the lower the risks of the project and the greater the chances that it will not disappoint investors' expectations. But this approach is very rough. Uncertainty and risk are relatively "subtle matters". This interpretation of NPV is like chopping thin sticks with an axe.

The biggest drawback of the DCF approach is that on its basis, it is not possible to develop a strategy for the development of the company in modern ones, which are characterised by three main properties: Complexity, Uncertainty and Ambiguity. Using the DCF approach in such an economic environment does not create strategic flexibility for the company. It does not allow the company management to improve the developed strategy, which is always associated with new investments, in accordance with the newly emerged circumstances. At the same time, it is crucial to predict these circumstances in advance without assuming that they will necessarily come, that is, to accept their probabilistic nature. This is the opportunity provided by the real options approach.

## 2. The Generalized Binomial Option Model

It would be stressed that we will deal only with investment projects that correspond to the company's development strategy. The ability to make new management decisions during the implementation of the investment project, contributing to the growth of NPV and the company's value as a whole, gives managers flexibility in managing investments in conditions of uncertainty. The cost of the flexibility of the investment project is equal to the value of all real options included in the project. The investment project, taking into account flexibility, is equal to the value of the NPV, increased by the value of the real options included in the project. So investment projects that have a negative NPV or close to zero receive a significant increase in value due to the ability to:

1) reduce, freeze or completely eliminate the negative processes that may begin during the implementation of the project;
2) develop the positive features of the project, spread its experience to other facilities;
3) postpone the project until new information that has commercial value is received;
4) change the corporate, investment or financial strategy in accordance with the new conditions of the external environment;
5) take advantage of new opportunities for financing projects and companies, and promptly change the structure and cost of capital.

We assume three classes of real options, divided into three types (see Fig.1.1). A particular investment project can contain one or more real options. In general, the more options an investment project contains the higher its value.

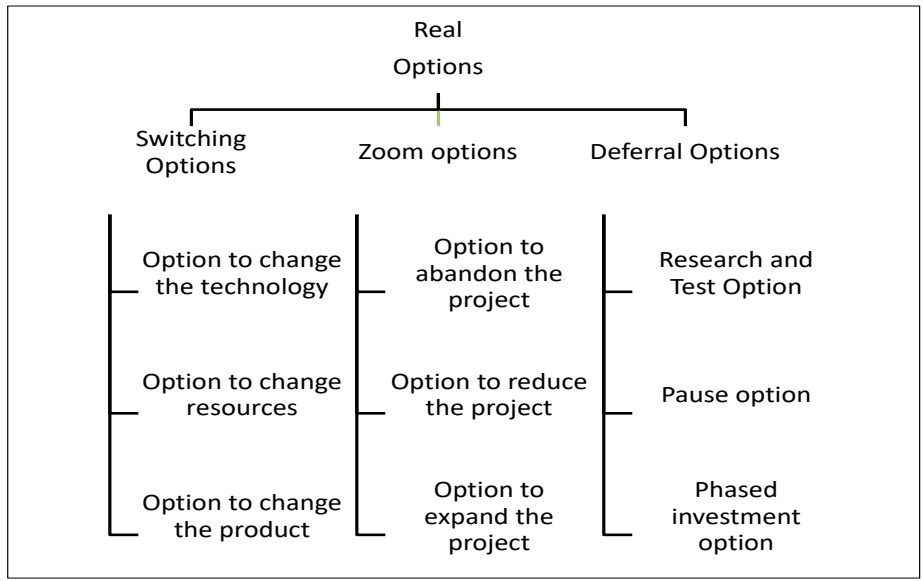

Fig. 2.1. Real Option Classification

The binomial model is a binary tree of well-defined outcomes. When building a binomial model, assumptions are used that (i) investors are neutral about risk, (ii) there can be only two scenarios simultaneously. The tree allows us to present all alternatives graphically and analytically to the development of events and, on their basis, make an informed decision. The option's value is calculated by moving from one point to another along the branches of the decision tree from left to right (Fig. 2.2).

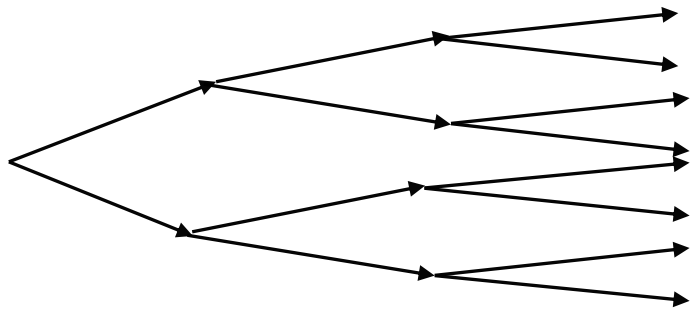

Fig. 2.2. Binomial Decision Tree

Firstly, we will present a generalized model of real options to assess the feasibility of launching an investment project (by investing $I_0$). We assume that in the process of implementing the project, there are opportunities to make changes to the project, which is interpreted as a real option. We proceed from the assumption that each such change is accompanied and caused by additional investment, which leads to new cash flows. After each change, we suppose the existence of two scenarios that can be interpreted as optimistic and pessimistic. We expect that changes will occur during the project's first two years. This presumption is justified by the assurance that, in actual practice, it is hardly possible to look into the future for more than two years. After two years of implementing the project with real options, further strategic development of the company can be considered by a new investment project, which will also be designed using real options.

To create the general model, we introduce two variables which control the real option process: $p$ as a probability of the real option and $\delta$ as a new investment, which in essence, creates the real option. We denote $\delta$ as an additional Cash Flow, which is negative if money is invested (corresponds to cash outflow) and positive, which corresponds to cash inflow. The case when $\delta < 0$ reflects the case when it is required to invest additional funds, which are strategically needed for creating a new opportunity for company development. The case when $\delta > 0$ can correspond, for example, to the opportunity to exit from the project by selling its assets and receiving money. In a trivial case, when $\delta = 0$, it assumes that all project assets remain unchangeable.

Fig 2.3 demonstrates the decision tree, which fills with details of the general scheme Fig. 2.2. according to these assumptions.

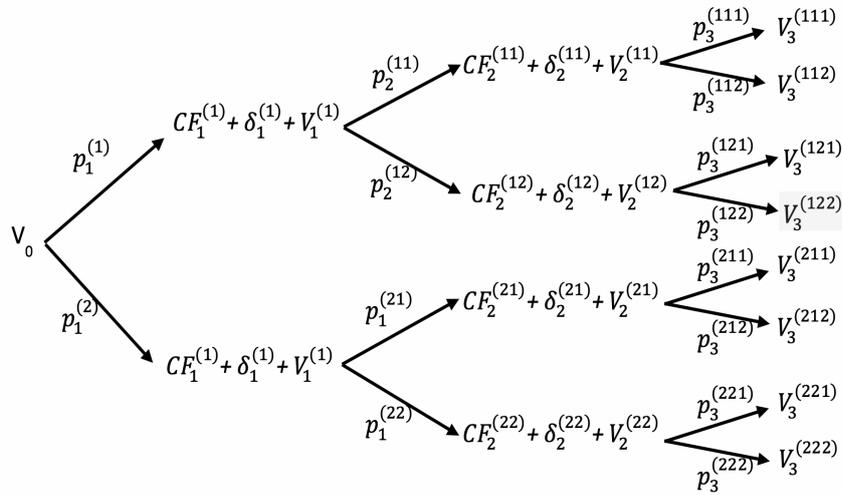

Fig. 2.3. Binomial Decision Tree for the Real Options Modelling

The scheme of Fig 2.3 allows us to build a step-by-step model for assessing project values. Here are the final equations for estimating the set of Project Values.

For the Project Value for time t = 0

$$V_0 = \frac{1}{(1+r)} \sum_{i=1}^{2} p_1^{(i)} \cdot V_0^{(i)}, \qquad (2.1)$$

where $p_1^{(i)}$ – probability of $i$-th option for 1-st year of the project, and $p_1^{(2)} = 1 - p_1^{(1)}$. $V_1^{(i)}$ – Project Value of $i$-th option for 1-st year of the project.

Project Values for time $t$ = 1:

$$V_1^{(i)} = \left(CF_1^{(i)} + \delta_1^{(i)}\right) + \frac{1}{(1+r)} \sum_{j=1}^{2} p_2^{(ij)} \cdot V_2^{(ij)}, i = 1,2, \qquad (2.2)$$

where $p_2^{(ij)}$ – probability of $j$-th option for the 2-nd year of the project for $i$-th option after 1-st year, $p_2^{(i1)} = 1 - p_2^{(i2)}$;

$CF_1^{(i)}$ – Cash Flow for $i$-th option for the 1-st year of the project;

$\delta_1^{(i)}$ – additional Cash Flow for i-th option for the 1-st year of the project as the result of possible making decision regarding the option; it can be $\delta_1^{(i)} < 0, \delta_1^{(i)} > 0, \delta_1^{(i)} = 0$;

$V_2^{(ij)}$ – Project Value of $ij$-th option for the 2-st year of the project.

Project Values for time t = 2.

$$V_2^{(ij)} = \left(CF_2^{(ij)} + \delta_2^{(ij)}\right) + \frac{1}{(1+r)} \sum_{l=1}^{2} p_3^{(ijl)} \cdot V_3^{(ijl)}, i, j = 1,2, \qquad (2.3)$$

where $p_3^{(ij)}$ – probability of $ij$-th option for the 3-rd year of the project, $p_3^{(ij1)} = 1 - p_3^{(ij2)}$;

$CF_2^{(ij)}$ – Cash Flow for $ij$-th option for the 2-nd year of the project;

$\delta_1^{(ij)}$ – additional Cash Flow for $ij$-th option for the 2-nd year of the project as the result of possible making decision regarding the option; it can be $\delta_1^{(ij)} < 0, \delta_1^{(ij)} > 0, \delta_1^{(ij)} = 0$;

$V_3^{(ijl)}$ – Project Value of $ijl$-th option for 3-rd year of the project.

Finally, Project Value for time t = 3.

$$V_3^{(ijl)} = \sum_{k=3}^{n} \frac{CF_k^{ijl}}{(1+r)^{k-3}}, i, j, l = 1,2. \qquad (2.4)$$

where $CF_k^{(ijl)}$ – Cash Flow for $ijl$-th option for the $k$-th year of the project.

After consecutive substitutions (2.4) to (2.3), then to (2.2) and finally to (2.1) we arrived at the final equation for the project value:

$$V_0 = \frac{1}{(1+r)} \sum_{i=1}^{2} p_1^{(i)} \left(CF_1^{(i)} + \delta_1^{(i)}\right) + \frac{1}{(1+r)^2} \sum_{l=1}^{2} \sum_{j=1}^{2} p_1^{(i)} p_2^{(ij)} \left(CF_2^{(ij)} + \delta_2^{(ij)}\right) +$$

$$+ \sum_{i=1}^{2} \sum_{j=1}^{2} \sum_{l=1}^{2} \sum_{k=3}^{n} \frac{p_1^{(i)} p_2^{(ij)} p_3^{(ijl)} CF_k^{(ijl)}}{(1+r)^k}. \qquad (2.5)$$

From the (2.5) we understand that to determine a real option sequence we need to assign the numbers for $p$ and $\delta$. For every sequence these variables should be assigned filling the table 2.1.

Table. 2.1. Control Variables for a real option sequence

| t = 1 | | | t = 2 | | |
|---|---|---|---|---|---|
| Scenario | p | δ | Scenario | p | δ |
| (1) | | | (1,1) | | |
| (2) | | | (1,2) | | |
| | | | (2,1) | | |
| | | | (2,2) | | |

Now, for clarity, we will demonstrate the features of the binomial model with the help of a specific example, choosing from a list of possible real options (see Fig. 2.1).

One way to increase the attractiveness of a project is to design the possibility of its reduction when the company begins to incur losses for some reason. In fairness, it is worth noting that this possibility is not always available. For example, suppose a company has signed long-term agreements that fix the volume of product supplies, purchases of raw materials, purchase prices and sales. In that case, there may be no opportunity to reduce the project, or the penalty amount will be so high that implementing the option will entail even greater losses. There are also organizational and technological obstacles to reducing the project.

Consider an example of estimating a reduction option using. Company SVP plans to start a project to produce plastic car parts. The volume of initial investments is $5,000K. Managers consider two scenarios with the same probability of realization of 0.5. The forecast of cash flows is presented in Table. 2.2. The company's cost of capital is estimated at 20%.

Table. 2.2. Cash Flow Forecast (without an option)

| year | 0 | 1 | 2 | 3 | 4 | Prob. |
|---|---|---|---|---|---|---|
| Scenario 1 | (5,000) | 2,000 | 2,400 | 2,600 | 3,500 | 0.5 |
| Scenario 2 | (5,000) | 1,000 | 1,200 | 1,300 | 2,000 | 0.5 |

Firstly, we can compute the project values using (1.3) and (1.4), as shown in Table 1.2.

Table. 2.2. Project Values (without an option)

| | Time | | |
|---|---|---|---|
| Project Value | 0 | 1 | Prob. |
| Scenario 1 | 4,955 | 7,831 | 0.5 |
| Scenario 2 | | 4,060 | 0.5 |

Therefore, the NPV of the investment project is

$$NPV = -\$5,000K + \$4,955K = -\$45K,$$

and the project must be rejected from the point of view of financial criteria.

We will try to make the project more attractive by designing a reduction option, which assumes that in the second (pessimistic) scenario, after the first year, it is possible to reduce the use of production capacity, which will lead to a decrease in current expenses and an increase in cash flows by 20% for each year. This event will require an additional investment of $ 100,000. The scheme of Fig 2.3 will be transferred to the simpler one (see Fig. 2.4) where $p_1^{(1)} = 0.5, \delta_1^{(1)} = 0, \delta_1^{(2)} = -\$100K$ and all the Cash Flows remains unchangeable.

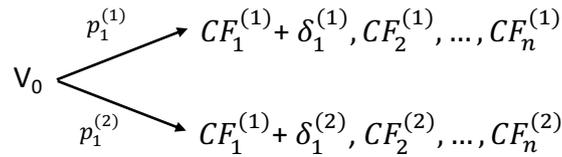

Fig. 2.4. Binomial Decision Tree for the reduction option

Cash Flows projection for the considered option is presented in Table 2.3.

Table. 2.3. Cash Flow Forecast (under the reduction option)

| year | 0 | 1 | 2 | 3 | 4 | Prob. |
|---|---|---|---|---|---|---|
| Scenario 1 | (5,000) | 2,000 | 2,400 | 2,600 | 3,500 | 0.50 |
| Scenario 2 | (5,000) | 900 | 1,440 | 1,560 | 2,400 | 0.50 |

Again, we can compute the project values using (2.1) and (2.2), as shown in Table 2.3.

Table. 2.3. Project Values (under the reduction option)

| | Year | | |
|---|---|---|---|
| Project Value | 0 | 1 | Prob. |
| Scenario 1 | 5,168 | 7,831 | 0.5 |
| Scenario 2 | | 4,572 | 0.5 |

Comparing the project value from Table 2.3 with the initial investment, we can figure out that under the reduction option, NPV becomes positive

$$NPV_{option} = -\$5,000K + \$5,168K = \$168K.$$

which means that the project can be accepted according to the financial criteria.

The option value can be defined as the difference:

$$V_{option} = NPV_{option} - NPV = \$168K - (-\$45K) = \$213K.$$

Based on the results of the calculations, it can be concluded that the reduction option makes the project attractive. The project value increased due to the possibility of reducing costs in the pessimistic version of the investment project.

Now consider a multi-stage Binomial model for the "Option of changing the product". Returning to our example, suppose that in the case of the pessimistic scenario, by the end of the first year, the company will be able to switch to the production of plastic parts for containers. The scheme of Fig 2.3 will be transferred to the simpler one (see Fig. 2.5), where the control variables and Cash Flows projection are presented in Table 2.5 and 2.6, respectively.

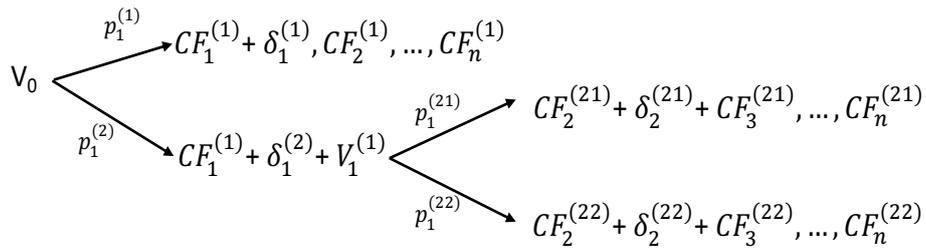

Fig. 2.5. Two-stage Binomial model with real option.

Table. 2.5. Control Variables (under the two options)

| Scenario | Prob. | Investment |
|---|---|---|
| Time = 1 | | |
| 1 | 0.5 | 0 |
| 2 | 0.5 | (100.00) |
| Time = 2 | | |
| 1,1 | 1.00 | 0 |
| 1,2 | 0.00 | 0 |
| 2,1 | 0.50 | (500.00) |
| 2,2 | 0.50 | (500.00) |

Table. 2.6. Cash Flow Forecast (under the switching option)

| year | 2 | 3 | 4 | Prob. |
|---|---|---|---|---|
| Scenario 1 | 940 | 3,200 | 4,200 | 0.50 |
| Scenario 2 | 940 | 2,200 | 3,500 | 0.50 |

Again, we can compute the project values using (2.2) and (2.3), as shown in Table 2.7.

Table. 2.7. Project Values (under the switching option)

| Year | | | | |
|---|---|---|---|---|
| 0 | 1 | | 2 | |
| 5,364 | 7,831 | 0.5 | | |
| | 5,043 | 0.5 | 5,583 | 0.5 |
| | | | 4,264 | 0.5 |

Based on the calculation results, it is concluded that the value of the project has increased and amounted to $5,364K. Given an initial investment of $5,000K, we conducted that NPV = $364K. To assess the option value, we remember that the Project's NPV with no options is negative $45K. Thus, the option value is computed as $364K - (-\$45K) = \$409K$.

The option for phased investment allows us to break down the investment project implementation into stages. The transition to the next stage is carried out only after the successful completion of the previous one. This is an effective tool for managing the risks of hazardous projects. The valuation of these options occurs in the same way as in the examples discussed above.

## 3. The Binomial-Random-Cash-Flow Real Option Model

An essential feature of the binomial model is the assumption that the values of cash flows are specified as numbers. At the same time, such a forecast is hardly possible in situations of the essential level of uncertainty for project implementation. Under such conditions, a decision-maker, at best, can forecast cash flows by assuming an error in that forecast. For example, he may expect cash flow by the end of the first year to be estimated at $240K with a margin of error of 10%. In other words, the cash flow forecast is assumed to be uncertain, and we need to develop a model for this uncertainty. Further, we will model cash flow uncertainty by using a probabilistic distribution which can be assigned in the framework of the subjective probability approach. This article proposes an Option Pricing Model that assumes projected cash flows are random variables with an assigned probability distribution. This assumption is considered to be a generalization of the binomial model. We call this model Binomial-Random-Cash-Flows (BRCF). Note that the random Cash Flows can be continuous or discrete.

We start studying BRCF with Gaussian probability distribution and call this model Binomial-Gaussian. Two reasons cause this distribution. First, it is a perfect way to model random variables largely affected by many factors. Secondly, in this case, it is possible to obtain final analytical assessments of the criteria for making decisions regarding the investment project. More generally, any probability distribution can be used. But then, for the decision-making procedure will use Monte Carlo simulation.

Fig 3.1 depictures the decision tree for the Binomial-Gaussian real options model.

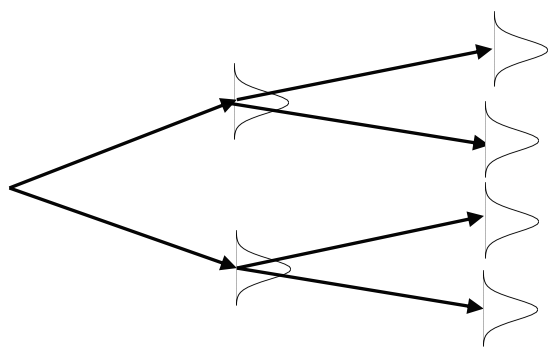

Fig. 3.1. Decision tree for Binomial-Gaussian model

The model's main assumptions are as follows:
1. Project scenarios are given by random variables of Cash Flows with Gaussian probability distributions.
2. The real option uses a binomial model with given probabilities and random outcomes for each decision tree point.

3. The real option's value is set considering the probability nature of the project value and based on the so-called Project Value at Risk (*PVaR*)[1], which is the maximum expected potential loss of the Project Value for a given confidence level α (0.01, 0.025, 0,05).
4. A (*1-* α) confidence boundary *PV*α of the random value of PV is a final estimate of the investment project feasibility:

$$\int_{PV_\alpha}^{\infty} f_{PV}(y)dy = 1 - \alpha, \tag{3.1}$$

where $f_{PV}(y)$ is a probability density function (pdf) of random variable PV. If *PV*α is less than the initial investment $I_0$, we can conclude that the project cannot be accepted as one that does not meet feasibility requirements.

The procedure for evaluating real options using the Binomial-Gaussian model contains the following steps:

1. Economic indicators of the basic version of the investment project are set, assuming the Gaussian distribution of random Cash Flows and the value of *PVaR* for the basic version is estimated.
2. Cash flows from the second year due to the real option are assigned by given means and standard deviations.
3. Considering the option's probability *p*, the PV's expected value and standard deviation are calculated, and *PVaR* for the project under the real option given is estimated.
4. If *the value of* $PV_\alpha = E(PV) - PVaR$ for the real option is higher than that in the basic version, then the option is recognized as attractive.

To demonstrate the Binomial-Gaussian technique, consider the simplest case when we have only one basic scenario for the investment project, which is given by Gaussian cash flows and a non-random discount rate *r*. According to the indicated procedure, find *PV*α for the basic scenario of the investment project. We assume that expected values $m_k = E(CF_k)$ and standard deviation $\sigma_k^2 = Var(CF_k) = E((CF_k - m_k)^2)$ of all the Cash Flows are assigned. Under these assumptions, we can find *PVaR* for the basic version of the project. Given PV as a linear function of cash flows, firstly, we find the Expected Value and the Variance of random PV:

$$m_{PV} = E(PV) = \sum_{k=0}^{n} \frac{m_k}{(1+r)^k}, \tag{3.2}$$

$$\sigma_{PV}^2 = Var(PV) = \sum_{k=0}^{n} \frac{\sigma_k^2}{(1+r)^{2k}}. \tag{3.3}$$

Then, given the Gaussian variable for *PV*, we can estimate *PV*α for the basic version of the project:

---

[1] The definition of PVaR is similar to the VaR in the theory of portfolio valuation.

$$PV_\alpha = m_{PV} - 1.64 \cdot \sigma_{PV}. \tag{3.4}$$

Let us now dwell on the situation when, based on the results of the project implementation, by the end of the first year decision is made on additional investments in $I_1$, which will lead to increased cash flows with a growth rate of $g$. We assign the probability of this real option as $p$. The problem is to assess $PV_\alpha$ for this scheme which can be presented in Fig. 3.2

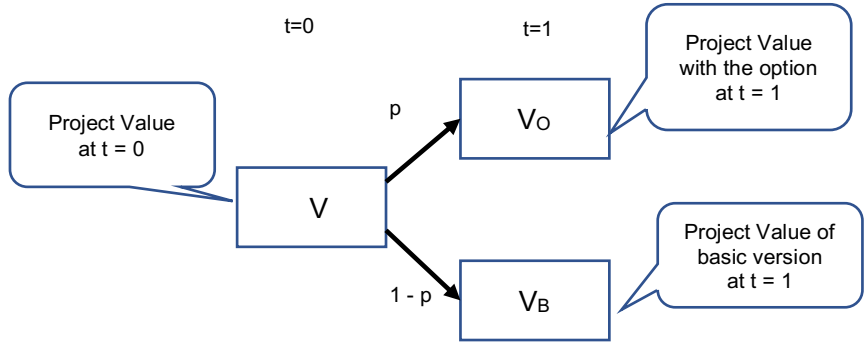

Fig. 3.2. Decision tree for the one-stage Binomial-Gaussian model

We estimate the Project Value at t = 0 as the discounted value of $V_O$ and $V_B$:

$$V = \frac{p \cdot V_O + (1-p) \cdot V_B}{(1+r)}, \tag{3.5}$$

where in turn, project values are equal to the discounted to the time t = 1 Cash Flows assigned according to the assumptions made:

$$V_B = \sum_{k=1}^{n} \frac{CF_k}{(1+r)^{k-1}}, \tag{3.6}$$

$$V_O = -I_1 + CF_1 + \sum_{k=2}^{n} \frac{CF_k \cdot (1+g)}{(1+r)^{k-1}}. \tag{3.7}$$

Note, that $I_1$ in (3.7) corresponds to the investment made to provide Cash Flow growth with the rate $g$ under the real option assigned. When we combine (3.5) – (3.7), we arrive at a final expression for the project value with the real option:

$$V = \frac{p \cdot (-I_1) + CF_1}{(1+r)} + \sum_{k=2}^{n} \frac{CF_k \cdot (1+p \cdot g)}{(1+r)^k}. \tag{3.8}$$

Again, we assume that Cash Flows in (3.8) are Gaussian random, and their expected values and standard deviation $\sigma_k$ are given. Since (3.8) is a linear function of $CF_k$, the project value $V$ is a Gaussian random variable, and we can find its mean value and standard deviation. To calculate the mean value $m_V$, we substitute in (3.8) the mean values of all $CF_k$ and $INV_1$, whereas the following formula can compute the standard deviation of the project value:

$$\sigma_V^2 = Var(V) = \frac{p^2 \cdot Var(I_1) + \sigma_1^2}{(1+r)^2} + \sum_{k=2}^{n} \frac{\sigma_k^2 \cdot (1 + p \cdot g)^2}{(1+r)^{2k}}. \qquad (3.9)$$

We can now estimate $PV_\alpha$ by using (3.4), for what we found $m_{PV}$ and $\sigma_{PV}$.

Now we will demonstrate the valuation of the real option according to the above algorithm. We return to the case considered in section 2. The company is going to start a project to produce plastic parts for cars. The forecast of cash flows, considering uncertainties, is presented in Table. 2.1. The value of the company's capital is assigned at 20%.

Table. 3.1. Cash flow forecast for the project

| year | 0 | 1 | 2 | 3 | 4 |
|---|---|---|---|---|---|
| Mean Value | (5,000) | 2,000 | 2,100 | 2,200 | 2,300 |
| Coefficient of Variation | - | 0.10 | 0.12 | 0.14 | 0.16 |

Using (3.1) and (3.2), we estimate $m_{PV} = 5{,}507$ and $\sigma_{PV} = 348$. Given these data, according to (3.3) we arrive at the $PV_{0.05} = 5{,}507 - 1.64 \cdot 348 = 4{,}936$. As the initial investment is 5,000, the basic version project does not look attractive as its NPV is negative.

Suppose now that, based on the results of the first year of the project implementation, the company management will be able to reduce current expenses, increasing cash flows by 20%. The volume of additional investments is estimated at $500K, with the standard deviation at $50K. The probability of the option is estimated at 50%. Using (3.6) and (3.7), we estimate $m_{PV} = 5{,}683$ and $\sigma_{PV} = 376$. Given these data, according to (3.3) we arrive at the $PV_{0.05} = 5{,}683.06 - 1.64 \cdot 376.59 = 5{,}065.46$.

Summarizing all the estimations in Table 3.2, we can decide that this real option can save the project, and the real option value can be assessed as 5,065 – 4,935 = 130.

Table. 3.2. Comparison of projects metrics

| Metrics/Alternative | Real Option Project | Basic Project's Version |
|---|---|---|
| PV Mean Value | 5,683 | 5,507 |
| PV Standard Deviation | 376 | 348 |
| PVaR | 617 | 572 |
| $PV_{0.05}$ | 5,065 | 4,935 |

The case discussed above, when Cash Flows were modelled using the Gaussian probability distribution, allows us to obtain estimates of the option's value and subsequent conclusions regarding the effectiveness of the real option in the final analytical form. This is often impossible to obtain such estimates for other probability distributions. In these cases, resorting to the Monte Carlo simulation is possible.

# 4. Monte Carlo Simulation of BRCF Real Option Model

The Monte Carlo techniques are universal tools regarding probability distributions and the complexity of the relationship between inputs and outcomes of the process under analysis. Hence, they are essential for uncertainty evaluation, and we use them for estimating the BRCF Option Pricing model.

Suppose we can create a BRCF model for a given real option scheme and present it in the form of the equation

$$Y = f(X_1, X_2, \ldots, X_n), \tag{4.1}$$

where $Y$ represents a random output quantity and $X_1, X_2, \ldots, X_n$ are the *n* random inputs.

In the particular case of the real option we considered in section 2, this equation was presented by (3.8) where $X_k = CF_k$ ($k = 1,2,\ldots,n$) and *Y* is a project value *V*.

With a set of generated samples, the distribution function for the value of the output *Y* in (4.1) will be numerically approximated by the following steps:

1. A set of random samples is generated from the probability density function (pdf) for each random input quantity $X_1, X_2, \ldots, X_n$. The sampling procedure is repeated *M* times for every input quantity.
2. The output quantities are calculated by:

$$y_i = f\left(x_1^{(i)}, x_2^{(i)}, \ldots, x_n^{(i)}\right) = f(\boldsymbol{x}^{(i)}),$$

where $i = 1,2,\ldots,M$.

3. Using the generated sample of $Y$, we can estimate pdf by a histogram, sample mean value and standard deviation.

Let us dwell on the Monte Carlo procedure for the real option we were dealing with in Section 3. Firstly, we will return to the Gaussian-Binomial model and compare the results of estimation obtained by Monte Carlo simulation with that from the previous case given by the final analytical form. Consider $CF_1, CF_2, \ldots, CF_n$ as independent Gaussian random variables with mean values and coefficients of variation from Table 3.1. For simulation, we used Oracle Crystal Ball software for the cases: 1) the basic version of the investment project without the real option and 2) the real option provided by additional investments, which resulted in the increase of Cash Flows by 20%. Fig. 4.1 demonstrates the pdf approximation obtained by the Monte Carlo software.

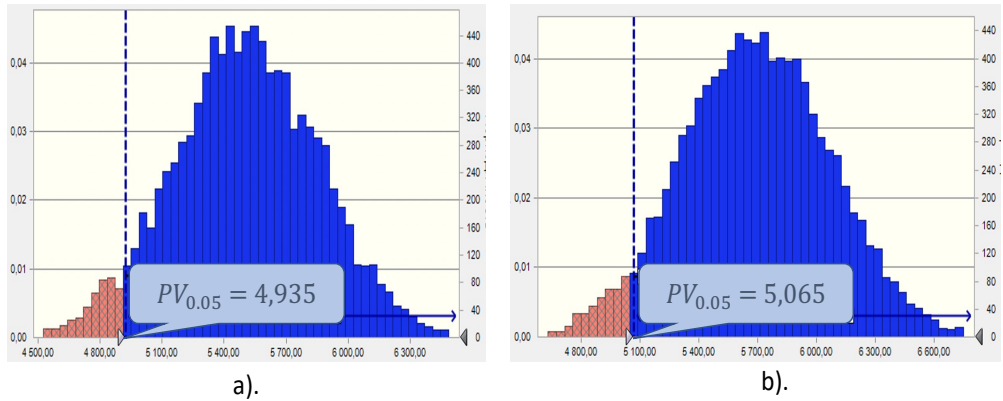

Fig. 4.1. Project Value pdf for the basic version (a) and the real option (b) (Gaussian CFs)

Table 4.1 summarizes statistics from the samples of the two Monte Carlo simulations.

Table. 4.1. Comparison of projects statistics by MC simulation

| Statistic/Alternative | Real Option Project | Basic Project's Version |
|---|---|---|
| PV Mean Value | 5,690 | 5,507 |
| PV Standard Deviation | 377 | 348 |
| *PVaR* | 626 | 572 |
| $PV_{0.05}$ | 5,065 | 4,935 |

Comparing the results of tables 4.1 and 3.2, we figure out that estimates obtained by the analytical and simulating approaches are almost identical.

Using the same procedure, we can assess the real option value for any pdf for cash flows. Let's do this for the uniformly distributed cash flows that correspond to the case when their uncertainties are assigned by the intervals, i.e. by two boundaries.

Initial data for the modelling are presented in Table 4.2.

Table. 4.2. Cash Flow forecast for the project (uniform pdf)

| CF Forecast / year | 1 | 2 | 3 | 4 |
|---|---|---|---|---|
| Left boundary | 1,800 | 1,848 | 1,892 | 1,932 |
| Right Boundary | 2,200 | 2,352 | 2,508 | 2,668 |

Fig. 4.2 demonstrates the pdf approximation obtained by the Monte Carlo software.

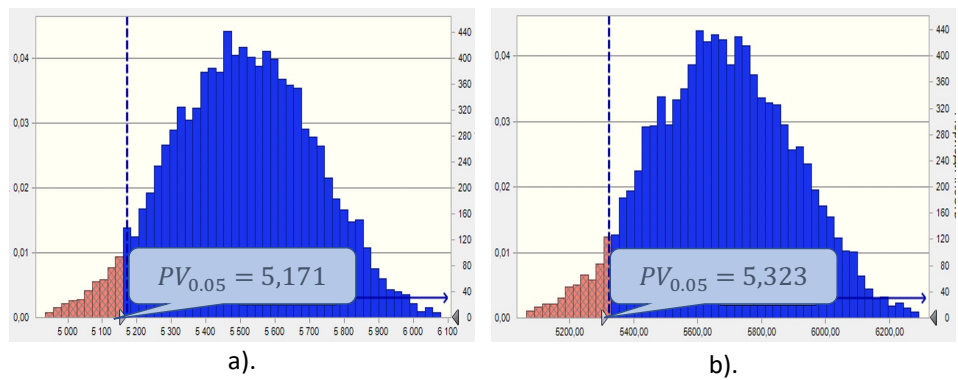

Fig. 4.1. Project Value pdf for the basic version (a) and the real option (b) (uniform CFs)

Table 4.3 summarizes statistics from the samples of the two Monte Carlo simulations for the uniform inputs.

Table. 4.3. Comparison of projects statistics by MC simulation (uniform CFs)

| Statistic/Alternative | Real Option Project | Basic Project's Version |
|---|---|---|
| *PV Mean Value* | 5,678 | 5,507 |
| *PV Standard Deviation* | 218 | 204 |
| *PVaR* | 355 | 336 |
| $PV_{0.05}$ | 5,323 | 5,171 |

As $PV_{0.05}$ is greater than the amount of investment, we can conclude that, in this case, all the versions should be recognized as feasible. But the Real Option project looks more attractive as it has more margin of safety.

The fact that we observe different conclusions for the two simulation cases is explained by the fact that the second case has less uncertainty for projected cash flows.

### *Conclusions*

In this paper, there were two primary objectives.

The first was offering to practice financial managers such models and computational algorithms that they could use to develop the company's strategy. Computational procedures have been proposed that make it possible to evaluate the Investment Project Value and thereby justify the feasibility of the strategic development proposed by the project manager for the case when changes to the project are made during the project's first two years. For this case, analytical equations are obtained that can be easily implemented by any calculation practice, for example, using Excel.

The second objective was related to the increasing level of uncertainty in the economic and business environment, which led to the assumption that the projected cash flows are random variables. The computational procedure resulting from achieving the first goal was developed for the case of Gaussian probability distributions of the projected cash flows. In this case, finite analytical equations were obtained that can be implemented using standard computing facilities. More generally, Monte Carlo simulation is recommended, requiring specialized software.

Each case was accompanied by practical examples that confirmed the validity of the proposed estimates and could demonstrate how the proposed models could be applied.